# Broadband polarization-independent low-crosstalk metasurface lens array-based mid wave infrared focal plane arrays

AYTEKIN OZDEMIR,[1,*] NAZMI YILMAZ,[2] FEHIM TAHA BAGCI[2] ,YUZURU TAKASHIMA[1], HAMZA KURT[2]

[1]College of Optical Sciences, University of Arizona, 1630 E. University Blvd., Tucson, AZ 85721, USA
[2] Department of Electrical and Electronics Engineering, TOBB University of Economics and Technology, Ankara 06560, Turkey
*Corresponding author: aytekin@optics.arizona.edu

**The miniaturization of pixel is essential for achieving high-resolution, planar, compact-size focal plane arrays (FPAs); however, the resulted increase in the optical crosstalk between adjacent pixels leads to serious drawback and trade-off. In the current work, we design and propose an efficient broadband polarization-insensitive all-dielectric metasurface lens array-based focal plane arrays (FPA) operating in the mid-wave infrared (MWIR). High focusing efficiency over 0.85 with superior optical crosstalk performance is achieved. We demonstrated that optical crosstalk can be reduced to low levels ≤2.8% with high efficiency. For the device performance, a similar figure-of-merit (FoM) from the previous reports was used and our device achieved FoM of 91 which outperformed all other types MWIR FPAs designed so far. Proposed metasurface lens arrays demonstrate great potential for increasing the signal to noise ratio and sensitivity thus paving the way for compact-size, high-resolution FPAs.**

*Metamaterials, Subwavelength structures, Micro-optical devices*

Metamaterials are artificial materials which are constructed from subwavelength structural components, and optical resonant effects of these subwavelength structural components allow metamaterials to show great potential in controlling electromagnetic properties via the electric permittivity and magnetic permeability [1,2]. Recently, metasurfaces, as the surface version or two-dimensional (2D) counterparts of metamaterials have been proposed as an alternative to the bulk three-dimensional metamaterials, due to their simple fabrication, lower losses, and smaller footprint [3]. Composed of two-dimensional arrays of subwavelength scatterers, metasurfaces show great capability to manipulate the amplitude, polarization, and phase of light in either transmission or reflection mode. Due to the superior control over the propagation of the electromagnetic waves, metasurface-based flat lenses [4-5], holograms [6] and wave plates [7] have been recently demonstrated. Mid-wave infrared (MWIR) focal plane arrays (FPAs) have important use in both civil and defense applications. To make compact-size, high-resolution, planar MWIR FPAs, an FPA with a smaller pixel pitch with high fill factor is necessary. However, reducing the pixel pitch size while keeping a large fill factor increases the spatial crosstalk between the adjacent pixels significantly as a strict and important trade-off. In other words, trend of reducing pixel size has some profound effects on pixel performance, in general, and optical performance since the wavelength of input light does not scale with technology and diffraction effects come into play. Furthermore, conventional MWIR FPAs require reduced optical aberrations, and this also results in a necessity of increasing the f-number, which increases the optical crosstalk as well [8]. For eliminating the drawbacks of decreasing pixel size, different methods, such as integration of microlens arrays with MWIR FPAs and mesa-isolation method were used [8]. Mesa-isolation method relies on the physical separation of the pixels, which require an etching process that can severely damage several pixels. As an example of mesa-isolation method, some previous studies on improving the performance of complementary metal oxide semiconductor (CMOS) image sensor pixels relied on placing light guides within pixels [9]. This enhanced the confinement of light inside each pixel by increasing the optical efficiency and prevented leakage out of the pixel by reducing optical crosstalk. Design of the investigated light guide mechanisms made use of total internal reflection (TIR) and reflection at metal-dielectric interface, respectively [9]. TIR based light guide design requires both a very high refractive index core material and a core region of high-aspect ratio etched over the central pixel area to increase the optical efficiency and decrease the optical crosstalk, but their fabrication is very challenging and inherent loss of metals limits the optical efficiency. Sub-wavelength photon-trapping (PT) structure-based pixel arrays were also shown as an effective means to improve the quantum efficiency and reduce the diffusion crosstalk [10]. However, PT structure had a slightly higher optical crosstalk compared to non-PT pixel arrays with similar geometry.

Relatively recent approach deploys microlens arrays to increase the optical fill factor of FPAs and CMOS image sensor pixels. They serve to focus and concentrate the incident light onto the pixel regions instead of allowing it to fall on non-photosensitive areas of the pixel array. Thus, integration of different types of microlens arrays to MWIR FPAs has been reported rigorously in a few studies so far [8,11,12]. Even if spherically refractive type microlens arrays reduced the focused spot size or the Airy disk, they suffered from the emergence of the first-order diffraction spots at the centers of the adjacent pixels and, thus, they were not able to enhance the optical crosstalk [8]. In another work, metallic metasurface lens arrays were shown to achieve an optical crosstalk of less than 1% but their focusing efficiency, which is defined as the ratio of the focused light power to the incident light power was very low (<11%) to make them practical in applications [11]. The primary reason for the poor efficiency performance of metallic metasurfaces is due to the intrinsic nonradiative Ohmic losses and cross-polarization scheme requirement. Because of these drawbacks of metallic metasurfaces, extensive research efforts have been performed based on all-dielectric metasurfaces [13-18]. Previous studies on all-dielectric metasurface-based lens array MWIR FPAs relied on Mie-scattering-based metasurfaces [12]. Even if these Mie-scattering-based metasurfaces demonstrated higher focusing efficiency and large figure-of-merit performance than their metallic counterparts, they operated successfully only in a very narrow spectral band and had an increased optical crosstalk due to poor sampling of the required lens phase profile. Instead of using the dipole resonance overlap at a fixed wavelength in Mie-resonance-based metasurfaces, we can use high aspect ratio silicon posts to generate truncated-waveguide modes that can be regarded as a non-resonant approach since the working principle is not dependent on a fixed wavelength [5,19]. These structures were also shown to have higher transmission and better phase sampling via the flexibility of lattice periodicity selection even for low f-number metalenses [5]. For implementation of the design, a metasurface platform based on high index contrast silicon posts was proposed. In addition, these high contrast dielectric meta-atoms are capable of achieving much higher transmission values comparing to the metallic metasurfaces and Mie-scattering-based metasurface counterparts. Thus, by integrating the metalens arrays designed with these meta-atoms to MWIR FPAs, high focusing efficiency can be achieved even with higher f-number values. In addition to that, by making use of these high aspect ratio posts, multipolar resonances, which allow for broadband operation, were ensured. To realize the abovementioned improvements on the performance of the MWIR FPAs, we used a metasurface platform with building blocks of metasurface composed of an array of amorphous silicon posts of different diameters which are resting on a sapphire substrate as shown in Fig. 1a. By using these all-dielectric meta-atoms, the metasurface lens arrays were designed, and the object signal and the noise during the detection of small objects with a low signal-to-noise ratio (SNR) was optimized to enhance the performance of the FPA device. The object signal was increased by maximizing the focused optical energy in the central pixel of the array via metasurface lens array. A figure-of-merit (FoM) previously defined was adapted such that ratio of the product of the f-number and the focusing efficiency to the optical crosstalk defines FoM [12]. More than 85% focusing efficiency with a sufficiently reduced optical crosstalk performance (≤ 2.8%) was achieved which outperform all previous MWIR FPA schemes. The designed high contrast dielectric metasurface FPAs achieved a FoM of 91, which is superior to all other forms of previously reported FPAs. In the next section, we present the design parameters of MWIR FPA and numerical results.

To design the constituent elements of the metalens arrays, transmission and resonance characteristics of the Si posts were analyzed. An array of a-Si posts of different diameters are arranged on a square lattice resting on a sapphire substrate as depicted in Fig. 1a. Each of the posts behave as a truncated waveguide with circular cross section supporting low quality factor Fabry-Pérot resonances [5,19,20]. The circular cross section of the posts results in a polarization-independent metasurface structure. Due to the high refractive index contrast between the posts and their surroundings, the posts have a weak optical coupling between them [6]. The weak optical coupling leads to an important consequence such that the phase coverage mainly depends on the diameter of posts and not on the distance between them. This allows the same local phase shift almost independent of the periodicity of the lattice by which metasurfaces are arranged. Figs. 1(b) and 1(c) show the simulated transmittance and phase of the transmission coefficient. The simulated transmittance and the phase of the amplitude transmission coefficient as only a function of the post diameter is also shown in Fig. 1(d). For modelling and simulating these Si posts, full-wave simulations were performed using the Lumerical finite-difference time-domain (FDTD) solver. For the MWIR band (3-to-5 μm) FPA design, wavelength of 3.2 μm, post height of 1.92 μm, lattice constant of 1.5 μm are considered. The simulation assumes periodic boundary conditions along the axial directions and perfectly matched layers (PML) boundary conditions were utilized in the perpendicular axis to Si posts. The refractive index data of a-Si was utilized from the data of Palik [21] and assumed as 3.435. The refractive index of sapphire was assumed as 1.70. Post height must be chosen such that the entire 0-2π phase range is covered by changing the post width, while keeping high transmission values. The lattice constant should be selected such that the lattice is non-diffracting. It should also satisfy Nyquist sampling criterion to supply sampling rate of required phase profile of metalens array. Fig. 1(d) shows the plot of one-to-one mapping between transmission phase and post diameters. To keep the transmission high, and to satisfy one-to-one relationship between the phase and post diameters, the diameter values corresponding to the sharp resonance were excluded from the design. Thus, the intensity transmission is kept above 91% while 0 to 2π phase range is covered. We also note that in contrast to Mie-scattering-based metasurfaces where only the electric and magnetic dipole resonances used, the resonant modes of the posts here contain higher-order electric and magnetic multipoles.

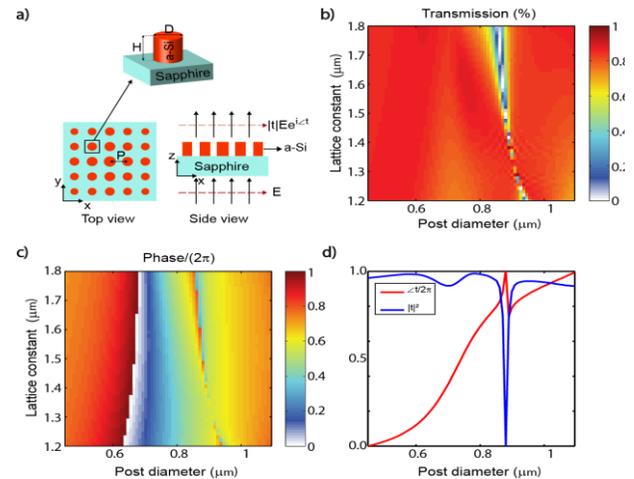

Fig. 1. (a) Schematic structure of metasurface with periodic a-Si circular posts resting on Sapphire substrate arranged in a square lattice (H=1.92 μm, P=1.50 μm), Top view (left), Side view (right). (b) Intensity transmission and (c) phase of the transmission coefficient variation as a function of a-Si post diameter (D) and period (P). (d) Simulated intensity and phase of the transmission coefficient for a fixed period P=1.50 μm at wavelength λ=3.2 μm with varied diameter. with varied diameter.

The required, ideal phase profile of a metalens can be given in Eq. (1)

$$\varphi(x,y) = \frac{2\pi}{\lambda}\left(f - \sqrt{f^2 + x^2 + y^2}\right). \quad (1)$$

To reduce the optical crosstalk while keeping the f-number as large as possible (such as 1.5), only the limited range of focal length values that provide a phase shift of minimum π radians between the edge and the center of a metalens in the lens array were utilized in the design. Therefore, our dielectric metasurface-based MWIR FPAs had focal lengths varying from 30 to 90 µm with aperture sizes varying from 20 to 30 µm considering the value of focal length. The required lens phase profile from Eq. 1, was realized by sampling it at the lattice sites by placing a scatterer from the periodic posts that most closely impart the desired phase change in transmission mode as seen in Fig. 2(a).

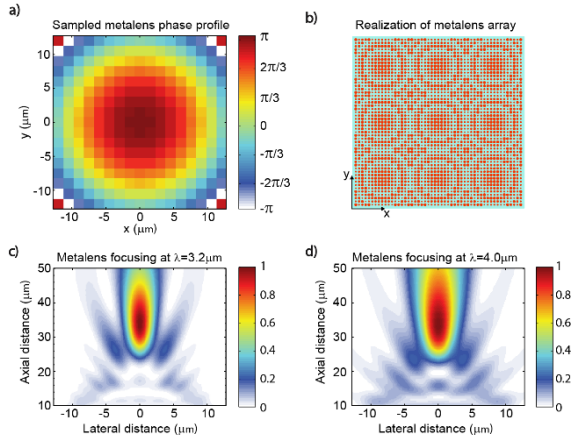

Fig. 2. (a) Sampled phase profile of a single metalens in the metalens array having a pitch length of 24 µm by arranging a-Si posts with unit cell size of 1.5 µm at the lattice sites. (b) Realization of metalens array by a-Si posts. (c) Focused light intensity distribution at the far-field for a wavelength of 3.2 µm with either TE or TM polarized light and (d) for another wavelength of 4 µm.

Next, metasurface lens array was realized with a discretized phase profile as demonstrated in Fig. 2(b). In Fig. 2(c), focused light intensity distribution at the far-field, scattered from the central metalens of the optimized design is shown. The operating wavelength is 3.2 µm. Comparing to the metallic metasurfaces that usually require the cross-polarization scheme to realize focusing, high contrast dielectric metasurface can focus both polarizations of light much more efficiently. As seen in Fig. 2(d), when source wavelength was changed to 4 µm, the focusing performance of the high contrast metasurface lens array outperformed the Mie-scattering-based metasurfaces, which are inherently narrowband.

Focused light intensity distribution at the far-field for different adjacent pixels of the optimized metalens array is also depicted in Figs. 3(a) and 3(b) at two different wavelengths of light. Furthermore, Fig. 3(c) and 3(d) show the cross-section of the focused light intensity and 3D view of the focused light intensity for the 3x3 metalens array at the focal plane, respectively. We should note that there is a sharp decay of intensity profile at each focal point. Besides, the peak intensity values are uniformly distributed at the array plane. Optical crosstalk of the FPA is related to the point-spread-function (PSF) [8]. Thus, optical crosstalk analysis of the proposed arrays can be realized by calculating the ratio of the corresponding PSF distributions inside the neighbor and the central pixels as defined by standard method, that was used before in the

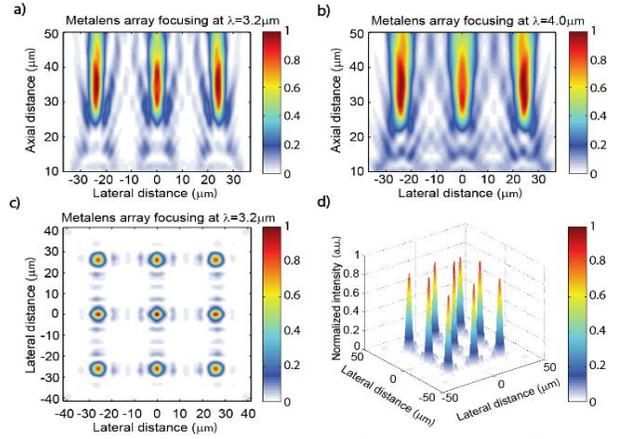

Fig. 3. (a) Far-field intensity distribution of light focused by different pixels of the optimized metalens array design at a wavelength of 3.2 µm and (b)at a wavelength of 4 µm. (c) and (d) show the cross-section of the focused light intensity and 3D view of the focused light intensity for the 3x3 metalens array at the focal plane, respectively.

other reported studies, which is given below by Eq. (2) [12]:

$$Optical\ Crosstalk = 100 \times \frac{\iint_{A_{n1eighbour}} PSF(x,y)dxdy}{\iint_{A_{central}} PSF(x,y)dxdy} \quad (2).$$

Optical crosstalk performances of the earlier designs were also compared to the proposed design here. Note that metallic metasurfaces have the lowest optical crosstalk (≤ 2%) for the f-numbers greater than 1.50. Since, they have the smallest unit cell dimensions and, thus, the best sampling of the required phase profile. Mie-scattering-type metasurfaces-based dielectric metalenses have been shown to provide lower optical crosstalk (≤ 3%) [12] than the conventional MWIR FPAs and the refractive microlenses; however, their optical crosstalk performance is worse than the metallic metasurfaces due to larger unit cell dimensions and, thus, poor sampling of the required phase profile. In contrast, the designed high index contrast dielectric metasurfaces gave rise to optical crosstalk value (≤ 2.8%) [11], which is better than Mie-type dielectric, but worse than metallic metasurfaces due to larger unit cell size again. High aspect ratio a-Si posts with smaller unit cell size also require a challenging fabrication process. If we compare focusing efficiency results, high contrast dielectric metasurfaces (≥85%) outperform the Mie-type dielectric metasurfaces (≥80%), and metallic metasurfaces (≤11%), therefore offer better practical implementation for metasurface lens arrays. Improved focusing efficiency can be attributed to the lack of polarization sensitivity and the elimination of the intrinsic absorption loss of the metallic metasurfaces [11,12]. In the design of conventional MWIR FPAs with a relatively lower optical crosstalk, smaller f-numbers are used; however, this leads to significant optical aberrations [8]. Considering the effect of all the aforementioned parameters, the overall FoM should be proportional to the f-number and the focusing efficiency while being inversely proportional to the optical crosstalk [12]. Thus, a FoM can be defined as the following:

$$FoM = \left[\frac{f/\#}{Optical\ Crosstalk}\right] \times \xi_{efficiency}. \quad (3)$$

Our dielectric metasurfaces were designed and optimized with a target of achieving the maxima of this FoM given by Eq. 3. Finally, FoM values of different metasurface-based MWIR FPAs are compared quantitatively in Fig. 4. Refractive microlens arrays do not improve optical crosstalk and the f-number due to the required for reduced crosstalk.

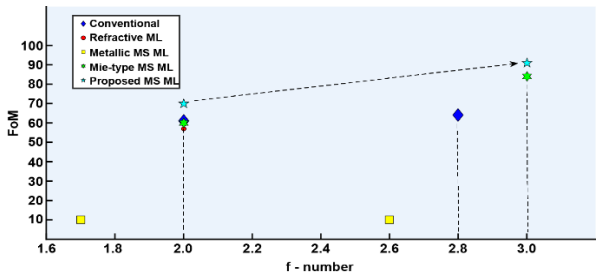

Fig. 4. FoM comparisons for different types of MWIR-FPA, illustrating the significantly improved performance of the proposed (waveguide-based) metasurface lensed FPAs (cyan pentagram marker) over the conventional FPAs (blue diamond marker degraded by higher optical crosstalk), the refractive microlensed FPAs (red circle marker degraded by diffraction noise), the metallic metasurface lensed FPAs (yellow square marker degraded by very poor transmission and focusing efficiency), Mie-type dielectric metasurface lensed FPAs (green hexagram marker suffering from higher optical crosstalk, lower focusing efficiency and also narrowband operation).

emergence of the first order diffraction spots and smaller f-numbers not improve optical crosstalk and the f-number due to the emergence of the first order diffraction spots and smaller f-numbers required for reduced crosstalk. Even if metallic metasurfaces have significantly reduced optical crosstalk, MWIR FPAs with metallic metasurfaces have the worst FoM due to their very poor focusing efficiency. As shown in Fig. 4, our proposed high index contrast dielectric metasurfaces have the highest FoM due to the achieved low optical crosstalk values with higher f-numbers, which cannot be possible with refractive microlens array and conventional MWIR FPAs. Remarkably, the increase of f-number also means that the imaging system can be made simple and lightweight Likewise, significant focusing efficiency improvement comparing to metallic metasurfaces allow our proposed design to be superior as well. The proposed metasurface structures can be fabricated at low cost using existing a single step conventional UV binary lithography. There might be some degraded FoM performance in experimental implementation of MWIR FPAs due to possible fabrication defects. For practical implementation of metasurface lens array-based FPAs, sensitivity to fabrication imperfections is numerically analyzed by considering random dimension errors caused by the electron beam lithography or reactive ion etching. Variation of FoM for different f-numbers and different random error variance of the constituent Si post diameters is shown in Fig. 5. For 60 nm fabrication error variance, FoM goes down by ≤ 64% for all the f-numbers. Besides, for fabrication error variance values ≤ 10 nm, designed metalens arrays have a good tolerance to fabrication error.

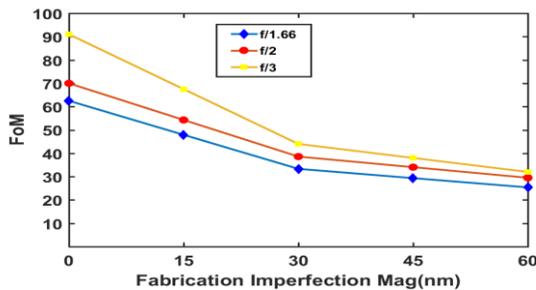

Fig. 5. Variation FoM as a function of the magnitude of the different fabrication imperfection variance values for different f-number of the metasurface lens elements in an array.

In conclusion, we have proposed and numerically demonstrated broadband, high-efficient and low-crosstalk dielectric metasurface lens arrayed MWIR FPAs by exploiting the unique properties of metasurfaces. Focusing efficiency and optical crosstalk performances of the designs were analyzed by implementing full-wave numerical simulations, and the achieved results were compared with performances of metalens arrays relying on metallic metasurfaces, Huygens' metasurfaces, refractive microlenses, and conventional FPAs. We obtained high focusing efficiency (≥85%) while achieving the best FoM value of 91 outperforming the FoM value of all other MWIR FPA types and keeping the optical crosstalk at a level of 2.8%. More importantly, designed metalenses perform much better than refractive microlens. The performance degradation of metasurface lens array due to manufacturing errors is also characterized and oblique illumination condition is investigated via inspecting the variations of the amplitude and phase of the transmitted light. These metasurfaces also pave the way for high spatial resolution, smaller pixel size and high SNR FPAs to achieve target detection and recognition of objects.

**Acknowledgement**

H.K. gratefully acknowledges the partial support of the Turkish Academy of Sciences.